\documentclass{aa}
\usepackage{graphicx}
\usepackage{natbib}
\usepackage{txfonts}
\begin{document}
\title{Modeling optical properties of cosmic dust grains using a distribution of hollow spheres}

\author{M. Min \and J.~W. Hovenier \and A. de Koter}

\institute{Astronomical institute Anton Pannekoek, University of Amsterdam, Kruislaan 403, 1098 SJ  Amsterdam, The Netherlands\\
\email{mmin@astro.uva.nl}}
\date{Received August 30, 2004, 2004; accepted November 10, 2004}

\abstract{
In this paper we study the combined effects of size and shape of small solid state particles on the absorption, emission and scattering characteristics. We use the statistical approach to calculate these optical properties. In this approach the average optical properties of an ensemble of particles in random orientation are represented by the average optical properties of an ensemble of simple shapes. The validity of this approach is studied in detail for a uniform distribution of hollow spheres where the fractional volume of the central inclusion is varied. 
We apply the results to two different areas of interest, i) infrared spectroscopy and ii) polarization of scattered light. The effects of particle size and shape on the optical characteristics are discussed. We compare the results using the distribution of hollow spheres with those obtained by using randomly oriented spheroids. Also we compare the results with observations and laboratory measurements. The distribution of hollow spheres is very successful in reproducing laboratory measurements of the scattering angle distribution of the degree of linear polarization for incident unpolarized light of randomly oriented irregular quartz particles. Furthermore, we show that we are able to derive the size distribution of dust grains by fitting the measured degree of linear polarization using computational result for hollow spheres. It is shown that the distribution of hollow spheres is a powerful tool for studying light scattering, absorption and emission by cosmic dust grains and in particular when large numbers of particle parameters need to be considered since the computational demand  of the distribution of hollow spheres is small.
}

\maketitle
%

\section{Introduction}
\label{sec:Introduction}

Dust is an important constituent of many astrophysical as well as atmospheric environments. The interpretation of observed infrared spectra, images and polarization requires knowledge of the scattering, absorption and emission properties of dust particles of various compositions, shapes and sizes. Optical properties calculated with the widely used Mie theory, which is valid for homogeneous spherical particles, often do not give satisfactory agreement with measurements and observations. Therefore, complicated methods have been developed to calculate the optical properties of irregularly shaped particles such as the Discrete Dipole Approximation (DDA) and the {T-matrix} method \citep[see e.g.][]{1973ApJ...186..705P, 1988ApJ...333..848D, MishHoveTravis, MishTravis2002, Wriedt2002, 2003JQSRT..79..775K}. For realistically shaped particles these methods are all very computationally expensive. Another approach to obtain optical properties of ensembles of irregularly shaped particles is the so-called statistical approach. In this approach the average optical properties of an ensemble of irregularly shaped particles are represented by the average properties of a distribution of simply shaped particles with a given shape distribution. Examples of simple shapes used in this way are spheroids and circular and hexagonal cylinders \citep{Mishchenko1994, Mishchenko1996b, 2002JQSRT..74..167K, 2004JQSRT..85..231K}. In \citet{Min2003b} it was shown that the statistical approach is very successful in calculating the absorption cross sections of an ensemble of randomly oriented very small particles, i.e. in the Rayleigh regime, as a function of wavelength. In that paper also results of calculations were presented for a uniform distribution of hollow spherical particles, varying the volume fraction occupied by the central inclusion. This shape distribution reproduces the measured absorption cross sections of small crystalline forsterite particles remarkably well. Because of their spherical symmetry the optical properties of hollow spheres can be obtained easily by using an extension of Mie theory. This makes this grain model an ideal candidate for studies where a large number of calculations for various grain sizes and/or compositions is required.

\citet{1988MNRAS.234..209J} used hollow spherical grains to model grain porosity in the ultraviolet to near infrared. The computational results using hollow spheres were compared with those obtained by using effective medium theories. The main motivation of that study was to find out if porous grains could enhance the interstellar extinction per unit mass. This turned out to be of much astrophysical significance in view of the so-called 'carbon crisis' which was brought up 7 years later by \citet{1995Sci...270.1455S}. It was found that, although by using porous particles the extinction is enhanced at some wavelengths, it is decreased at other wavelengths, consistent with the expectation from the Kramers-Kronig relations \citep{1969ApJ...158..433P}. However, \citet{1988MNRAS.234..209J} argues that this problem might be solved by changing the size distribution of interstellar grains. We will discuss the effect of grain shape and porosity on the interstellar silicate extinction feature in the mid infrared by applying the hollow sphere model.

In this paper we will use a distribution of hollow spheres to calculate absorption, extinction and scattering properties for various particle sizes beyond the Rayleigh regime. We compare our results to those obtained by using a shape distribution of spheroidal particles and to laboratory measurements of the scattering properties of irregular particles.

In Sect. \ref{sec:Dust shape models} we introduce the dust shape models employed in this paper. Then we will concentrate on two main areas; i) absorption and extinction spectra (Sect. \ref{sec:The absorption, extinction and emission spectra}) and ii) the degree of linear polarization (Sect. \ref{sec:The degree of linear polarization}). Some examples of applications in both areas are provided. In Sect. \ref{sec:Discussion} we discuss the physical interpretation of the results. 

\section{Dust shape models}
\label{sec:Dust shape models}

In order to understand the effects of particle shape and internal structure on the emissivities of dust grains we consider two simple dust models. As a first approximation we change the internal structure of homogeneous spherical dust grains in the most simple way by using a distribution of hollow spherical particles, varying the volume fraction occupied by the central inclusion. In the second distribution we change the external shape of the homogeneous spherical particles by using a distribution of spheroidal particles, varying the aspect ratio. In these shape distributions the material volume of the particles is kept constant. Both shape distributions have been studied in the Rayleigh limit, i.e. when the particles are very small compared to the wavelength inside and outside the particle. In \citet{Min2003b} it was found that in this limit the absorption properties for the distribution of spheroidal particles and for the distribution of hollow spherical particles are very similar and that both shape distributions provide good agreement with laboratory measurements of the absorption cross sections as a function of wavelength in the infrared part of the spectrum. This suggests that the exact shape of the particles is to a certain degree not important as long as we break the perfect symmetry of a homogeneous sphere. When applying the statistical approach one needs to realize that the particle shapes used contain only very limited information on the real shapes of the dust grains and that additional research has to be done in order to decode this information into physical characteristics of the dust grains \citep[see also e.g.][]{2002JQSRT..74..167K}.

\subsection{Distribution of Hollow Spheres (DHS)}
\label{sec:Distribution of Hollow Spheres (DHS)}

The simplest internal deviation from a homogeneous sphere is a hollow spherical shell with the same material volume. In \citet{Min2003b} a Distribution of Hollow Spheres (DHS) is used to calculate the absorption cross sections for very small particles. In this shape distribution we simply average over the fraction $f$ occupied by the central vacuum inclusion in the particle from zero to some value ${f_\mathrm{max}\leq 1}$, giving equal weights to all values of $f$ while the material volume of the particle is kept the same. This means that the particles with higher values of $f$ have a larger outer radius. 
Thus, the distribution is given by
\begin{equation}
\label{eq:shape dis}
n(f)=\left\{ \begin{array}{ll}
1/f_\mathrm{max}, & 0\leq f< f_\mathrm{max}, \\ 
 & \\
0, & f\geq f_\mathrm{max}, \end{array}\right.
\end{equation}
where $n(f)df$ is the fraction of the number of particles in the distribution between $f$ and $f+df$. 
Unless stated otherwise, in this paper we will take ${f_\mathrm{max}=1}$, the most extreme shape distribution. 
We have to note here that particles with $f=1$ and a finite volume have an infinitely large outer radius. For particles much smaller than the wavelength, integrating the optical properties up to $f=1$ can be done analytically \citep{Min2003b}. For larger particles numerical computations are necessary, but these are not possible for $f=1$. However, we then chose $f_\mathrm{max}$ large enough to reach convergence to the values for $f_\mathrm{max}=1$. For most cases we considered, it suffices to integrate up to $f=0.98$. The choice of the shape distribution we take is motivated by the fact that for many applications one of the parameters of interest is the dust mass. Therefore, it is convenient to choose a shape distribution in which the total mass of the particles is conserved. Other choices of the shape distribution, e.g. conserving the surface area of the particles, might be convenient in other cases.
The equations for the optical cross sections in the Rayleigh limit are provided by \citet{vandeHulst}. In \citet{Aden1951} the general equations for light scattering and absorption by layered spheres of arbitrary size are derived as a simple extension of Mie theory \citep{Mie}. By taking the refractive index of the core of a layered particle to be that of vacuum these general equations can be used to calculate the optical properties of hollow spherical particles for almost all particle sizes.

The calculations were done using the layered sphere code for which the basic ideas are explained in \citet{1981ApOpt..20.3657T}. We successfully performed the Cloude test \citep[see][]{Hovenier2000} on the $4\times 4$ scattering matrix obtained with this code. Also we checked that the computed values of the scattering matrix obtained for forward and backward scattering are in agreement with theoretical values derived from symmetry arguments. In addition, we considered the limiting case $f\rightarrow 0$ and compared this with calculations for homogeneous spheres, all with satisfying results.

\begin{figure*}[!htbp]
\begin{center}
\resizebox{\hsize}{!}{\includegraphics{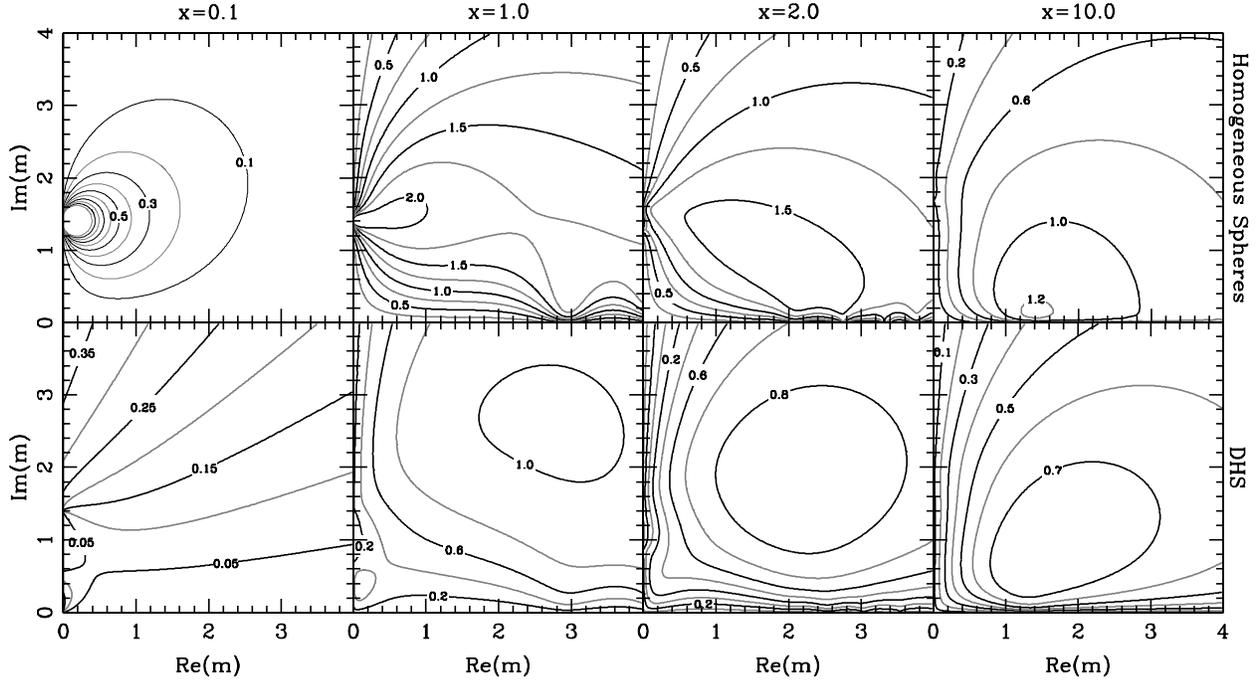}}
\end{center}
\caption{The shape averaged absorption efficiency as a function of the real and imaginary part of the refractive index for various values of the dimensionless size parameter ${x=2\pi r_V/\lambda}$, where $r_V$ is the radius of a volume equivalent sphere and $\lambda$ is the wavelength of incident radiation. The upper panels are for homogeneous spheres, the lower panels for the distribution of hollow spheres (DHS).}
\label{fig:Contours}
\end{figure*}

\subsection{Uniform Distribution of Spheroids (UDS)}
\label{sec:Uniform Distribution of Spheroids (UDS)}

The simplest external deviation from a homogeneous sphere is a homogeneous spheroid with the same material volume. In \citet{Min2003a} a method was developed to calculate the optical cross sections of a Uniform Distribution of Spheroids (UDS) for every particle size using a combination of the {T-matrix} method \citep{Mishchenko1996a} and some approximate methods. In the UDS we average over the aspect ratios (i.e. the ratio between the major and minor axis) of the spheroids giving equal weights to all oblate and prolate spheroidal shapes. The material volume of all particles in the shape distribution is the same. For details we refer to \citet{Min2003b, Min2003a}. Although the average absorption cross sections for this distribution can be calculated, the computations become very slow for intermediate sized particles (with sizes comparable to the wavelength of radiation) where we still have to apply the {T-matrix} method.

In this paper we will average the spheroidal shapes over all prolate and oblate shapes up to an aspect ratio of $10$. This is at the moment the most extreme shape distribution for which calculations can be done in a reasonable time. For the {T-matrix} calculations we used the well tested code developed by \citet{Mishchenko1998a}.

\section{The absorption, extinction and emission spectra}
\label{sec:The absorption, extinction and emission spectra}

Absorption, extinction and emission spectra of dust provide crucial information on the size and composition of the dust grains. The thermal emission spectrum of a dust grain in thermal equilibrium averaged over all orientations is proportional to the orientation averaged emission cross section times a Planck function. From Kirchhoff's law we know that this emission cross section is equal to the orientation averaged absorption cross section. The size of a particle, $r_V$, is defined to be the radius of a sphere with the same material volume as the particle. We define the average absorption cross section, ${\left<C_\mathrm{abs}\right>}$, to be the shape- and orientation-averaged absorption cross section of a collection of particles with a shape distribution  and a fixed value of the material volume equivalent radius. In this paper we will always first average over all particle orientations. After this we will average over particle shape, which is denoted by one bracket ${\left<..\right>}$. For calculations of the optical characteristics of a particle it is useful to define the dimensionless size parameter
\begin{equation}
x=\frac{2\pi r_V}{\lambda},
\end{equation}
where $\lambda$ is the wavelength of incident radiation. The mass absorption coefficient $\kappa$ is the average absorption cross section per unit mass
\begin{equation}
\kappa=\frac{\left<C_\mathrm{abs}\right>}{M}.
\end{equation}
Here $M$ is the mass of the particles in the collection with the same $r_V$.

The orientation averaged absorption efficiency of a particle defined as ${Q_\mathrm{abs}=C_\mathrm{abs}/G}$, with $G$ the orientation averaged geometrical shadow, only depends on the shape, size parameter and complex refractive index $m$ of the particle. Because the distribution of normals to the surface of all convex particles is uniform \citep{vandeHulst}, the absorption efficiency of very large convex solid particles averaged over all orientations is the same as for spheres and can be written as \citep{BohrenHuffman}
\begin{equation}
\label{eq:large particle abs}
Q_\mathrm{abs}=1-\mathcal{R},
\end{equation}
where $\mathcal{R}$ is a reflection factor which does not depend on the size and shape of the particle but only on the complex refractive index $m$. Eq.~(\ref{eq:large particle abs}) also holds for particles with one or more inclusions when the outer surface of the particle is convex (e.g. hollow spheres). Therefore, for very large particles the absorption efficiency and cross section averaged over all orientations as a function of wavelength will have the same behavior for all particles with a convex outer surface.

In Fig.~\ref{fig:Contours} the shape averaged absorption efficiency, ${\left<Q_\mathrm{abs}\right>}$, is plotted as a function of the real and imaginary part of the refractive index for various values of the size parameter for homogeneous spheres (upper panels) and for the distribution of hollow spheres (DHS, lower panels). We clearly see that there is a large difference between the contours for homogeneous spheres and those for the DHS when $x$ is small and that this difference slightly disappears when we go to larger values of $x$, which is to be expected from Eq.~(\ref{eq:large particle abs}). 
The difference between the two types of contours is already small for ${x=10}$, which is much too small to expect Eq.~(\ref{eq:large particle abs}) to be valid. Based on the assumptions used to derive Eq.~(\ref{eq:large particle abs}), it is expected to be valid only when ${x\gtrsim 60}$. This suggests that for particles with a size parameter larger than about $10$ with a convex outer surface the behavior of the absorption cross section as a function of wavelength is to a certain degree independent of particle shape, which allows us to use homogeneous spheres to calculate the absorption cross sections for $x\gtrsim 10$.

\subsection{Example: amorphous silicates}
\label{sec:Example: amorphous silicates}

\begin{figure}[!tbp]
\resizebox{\hsize}{!}{\includegraphics{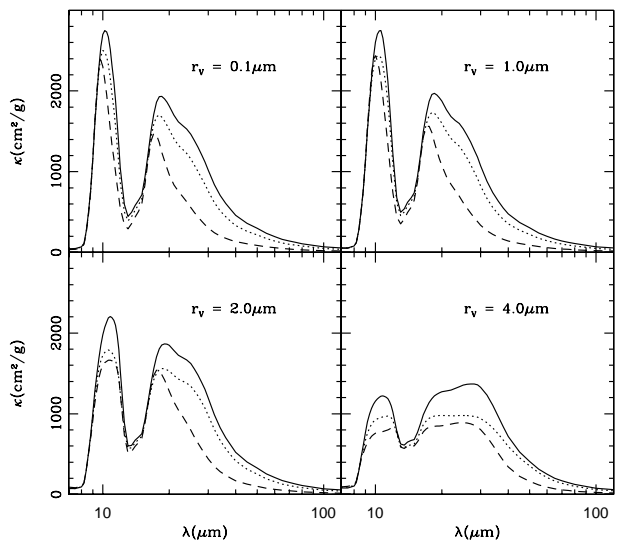}}
\caption{Absorption cross section per unit mass for amorphous olivine for different shape distributions and grain sizes. The solid line is for the distribution of hollow spheres (DHS), the dotted line for the uniform distribution of spheroids (UDS) and the dashed line for homogeneous spheres.}
\label{fig:Amorphous}
\end{figure}

\begin{figure}[!tbp]
\resizebox{\hsize}{!}{\includegraphics{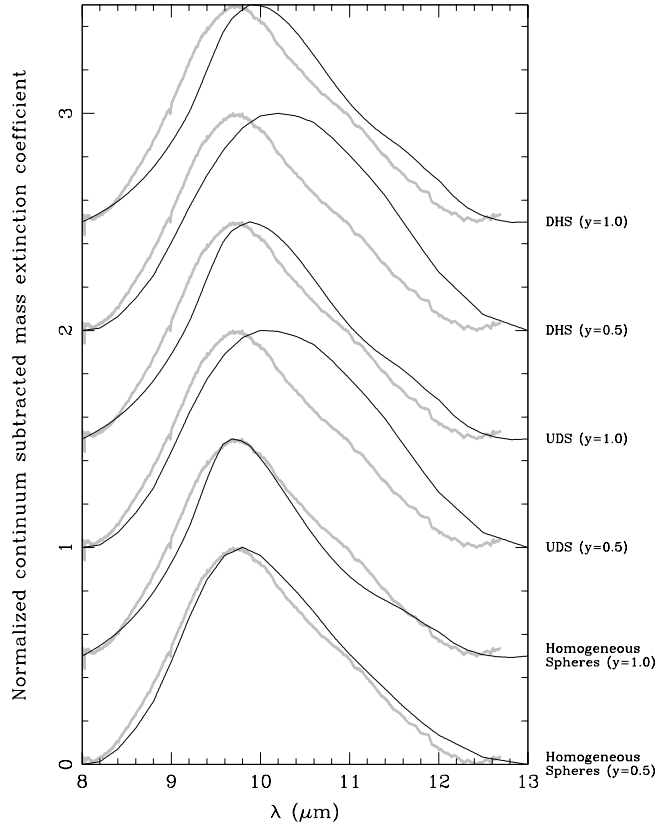}}
\caption{The normalized, continuum-subtracted mass extinction coefficients for various grain shape distributions and chemical compositions (black curves). The grain size $r_V=0.1\,\mu$m. Also plotted is the normalized, continuum subtracted extinction of interstellar dust grains as observed towards the Galactic center taken from \citet{Kemper2004} (gray curves). The refractive indices as functions of wavelength are taken from \citet{2003A&A...408..193J} for the $y=1.0$ data and from \citet{1995A&A...300..503D} for the $y=0.5$ data. The continuum that was subtracted from all spectra is taken to be a straight line from $8$ to $13\,\mu$m. The normalization is done such that the maximum value of all curves equals unity. The curves are shifted with respect to each other with intervals of $0.5$.}
\label{fig:Galaxy}
\end{figure}

As a first example we consider the infrared mass absorption coefficient of amorphous olivine (Mg$_{2y}$Fe$_{2-2y}$SiO$_4$ with ${0\le y\le 1}$) which is one of the most abundant species found in astronomical environments. This silicate is observed by looking at the thermal emission from, for example, comets \citep[see e.g.][]{Crovisier1997, 1994ApJ...425..274H}, proto-planetary disks \citep[see e.g.][]{Bouwman2001b, 2003A&A...400L..21V}, asymptotic giant branch stars \citep[see e.g.][]{1999A&A...350..163M, 2002A&A...389..547H} as well as by looking at the extinction by interstellar dust grains \citep[see e.g.][]{1984ApJ...285...89D, 2001ApJ...550L.213L, Kemper2004}. The exact behavior of the mass absorption and extinction coefficients as functions of the wavelength for various grain sizes is therefore crucial in understanding the formation and processing of grains in these environments. The observed spectrum is dominated by two broad features, one at $9.8\,\mu$m and one at $17.5\,\mu$m. The $9.8\,\mu$m feature has been studied very extensively because the atmospheric window around $10\,\mu$m allows ground based observations.

Fig~\ref{fig:Amorphous} shows the mass absorption coefficient, $\kappa$, as a function of wavelength for amorphous olivine grains with various sizes and $y=0.5$. The refractive indices as a function of wavelength are taken from \citet{1995A&A...300..503D, 2003A&A...408..193J}. We see that for the smaller grain sizes, the $17.5\,\mu$m feature broadens significantly when going from homogeneous spheres towards the distribution of hollow or spheroidal particles. This becomes important when identifying features in the region $20-30\,\mu$m.

It should be noted that all spectroscopic features that we considered are broadened with respect to those calculated using homogeneous spherical particles. Actually, it might be argued that the correct way to put it is that the features as calculated using homogeneous spheres are sharpened with respect to those of all other particle shapes. The reason for this is that small homogeneous spherical particles exhibit a rather artificial sharp resonance close to when the refractive index $m=\sqrt{-2}$. For irregular particles the mass absorption coefficient is a smoother function of the refractive index which has its maximum approximately when the bulk refractive index is at its maximum. This causes the difference in feature shape observed in, for example, the 18 micron feature in Fig.~\ref{fig:Amorphous}.

Many sources show a $9.8\,\mu$m extinction or emission feature that is remarkably compatible with amorphous olivine grains that are very small ($r_V\lesssim 0.1\,\mu$m), homogeneous in composition, spherical in shape and have a magnesium/iron content with ${y=0.5}$. When we change the shape of the particles, the $9.8\,\mu$m feature shifts towards longer wavelengths and broadens (see Fig.~\ref{fig:Amorphous}) which is not compatible with observations. This is puzzling since for example images of interplanetary dust particles show that these grains have a very irregular shape \citep{DustCatalog}. In addition grains in interstellar space produce observable linear polarization in the forward scattering direction which can only be explained using non-spherical dust grains that are aligned to a certain degree \citep[see e.g.][]{1995ApJ...450..663H}. Fig.~\ref{fig:Galaxy} shows the normalized continuum subtracted mass extinction coefficient in the $10\,\mu$m region of amorphous olivine grains with ${y=0.5}$ and ${y=1.0}$. Note that here we plot the mass \emph{extinction} coefficient, which is just the average extinction cross section per unit mass. For very small particles the scattering is negligible, in which case the mass extinction coefficient is equal to the mass absorption coefficient, $\kappa$. The figure shows the normalized continuum subtracted mass extinction coefficient for small homogeneous spherical particles, for the uniform distribution of spheroids (UDS) and for the distribution of hollow spheres (DHS) all with $r_V=0.1\,\mu$m. The continuum that was subtracted is taken to be a straight line between $8$ and $13\,\mu$m. The normalization of all curves in Fig.~\ref{fig:Galaxy} is such that the maximum value equals unity. Also plotted in this figure is the normalized continuum subtracted extinction towards the Galactic center \citep{Kemper2004}. We see that the extinction spectrum towards the galactic center is fairly well reproduced by both curves with homogeneous spherical particles. The feature obtained by using the spheroids or hollow spheres with $y=0.5$ is much broader than the galactic extinction feature. Also the maximum for these grains is shifted towards longer wavelengths than observed. For the magnesium silicates (${y=1.0}$), the agreement with the interstellar extinction feature for the DHS and UDS shape distributions is better. Because of the observed degree of linear polarization the interstellar grains are most likely non-spherical \citep[see e.g.][]{1996A&A...309..258G}. A possible explanation for the shape of the observed interstellar extinction feature might be that the grains in the interstellar medium have a very high magnesium over iron ratio. Note that the particle ensembles considered here are in random orientation and would therefore not produce any linear polarization in the forward scattering direction. However, aligned, non-spherical dust grains produce an extinction spectrum which is more similar to that of randomly oriented non-spherical or non-homogeneous grains than to that of homogeneous spheres.

\subsection{Example: crystalline silicates}
\label{sec:Example: crystalline silicates}

\begin{figure}[!tbp]
\resizebox{\hsize}{!}{\includegraphics{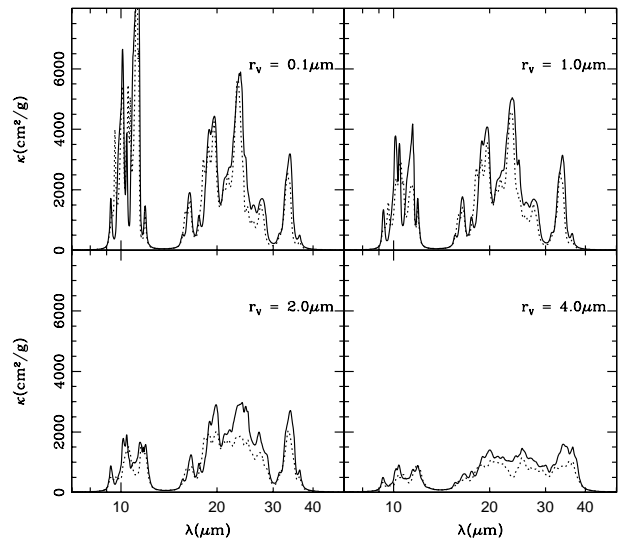}}
\caption{Mass absorption coefficient, $\kappa$, for various sizes of crystalline forsterite particles. The solid line is for the distribution of hollow spheres (DHS), the dotted line is for the uniform distribution of spheroids (UDS). The refractive index as a function of wavelength is taken from \citet{Servoin}.}
\label{fig:Crystalline}
\end{figure}

Another example of the importance of grain shape for the absorption and extinction spectra are the spectra of crystalline silicate particles. Small crystalline silicate grains have a very clear spectral signature in the infrared part of the spectrum which makes them easy to detect in for example thermal emission spectra \citep[e.g.][and references at the start of Sect. \ref{sec:Example: amorphous silicates}]{Bouwman2003}. Crystalline silicates can form only under certain conditions from gas phase condensation or by annealing of amorphous silicates. The abundance of crystalline silicates is therefore a good probe for the processing history of the dust grains.   

The mass absorption coefficient as a function of wavelength for small crystalline silicate grains shows many very narrow and strong resonances in the infrared part of the spectrum. Because of the strength of the resonances, their shape and wavelength positions are very sensitive to the particle shape. For example, calculations of the mass absorption coefficient as a function of wavelength using homogeneous spherical particles completely fail to reproduce laboratory measurements on small crystalline forsterite grains \citep{Fabian, Min2003b}. A very successful shape model which is frequently used to calculate the absorption spectra of these grains is the so-called Continuous Distribution of Ellipsoids (CDE) \citep{BohrenHuffman}. This distribution takes all possible ellipsoidal shapes into account. Although the measured absorption spectrum is very well reproduced using this shape distribution, a major disadvantage is that calculations for this distribution are restricted to the Rayleigh domain since it uses ellipsoidal shapes, and it is thus not suitable for studying combined size/shape effects. We performed calculations using the shape distribution suggested by \citet{2002MNRAS.334..840L}. However, the results for, for example, the wavelength positions of the crystalline silicate resonances were poorly reproduced, in contrast with using CDE. The shape distributions we use here also reproduce the measurements on small forsterite grains \citep{Min2003b} and can be used outside the Rayleigh domain.

\begin{figure}[!tbp]
\resizebox{\hsize}{!}{\includegraphics{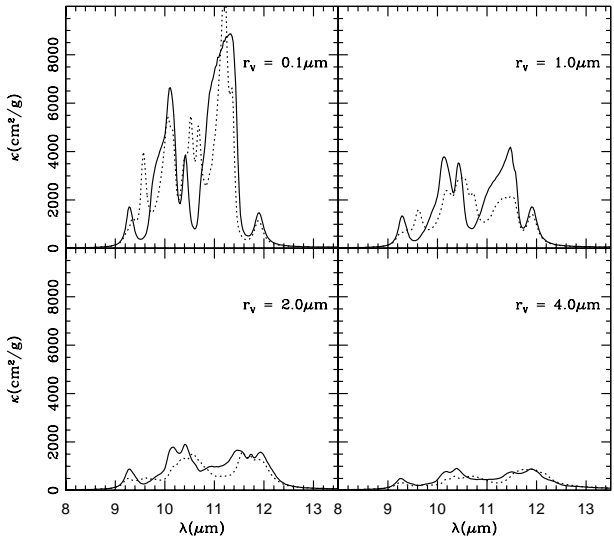}}
\caption{Same as Fig.~\ref{fig:Crystalline} but only for {$\lambda=8-13.5\,\mu$m}.}
\label{fig:Crystalline 10 micron}
\end{figure}

Fig.~\ref{fig:Crystalline} shows the mass absorption coefficient as a function of wavelength for various grain sizes. The refractive index as a function of wavelength is taken from \citet{Servoin}. The solid curve is for the distribution of hollow spherical particles (DHS), the dotted curve is for the uniform distribution of spheroids (UDS). We see that the strong resonances visible for very small particles disappear when we increase the particle size. We notice that the resonances at shorter wavelengths disappear first. Also, the stronger resonances disappear before the weaker resonances do. These effects are due to the fact that the important parameter which determines the absorption properties is ${|2\pi~m~r_V/\lambda|}$. So the strength of the resonances is influenced by the wavelength, the refractive index and the size of the particle. This effect causes the ratios between the various resonances to change when we increase the particle size. In the interpretation of observations this ratio can be used as an indicator of the typical particle size.

We now compare the curves calculated for the uniform distribution of spheroids with those calculated for the distribution of hollow spheres in Fig.~\ref{fig:Crystalline}. First we note that for very small particles ($r_V=0.1\,\mu$m) the differences are very small. When the grain size is increased the trends seen in the two curves are the same, as is described above. However, we see that the resonances for a given grain size obtained from the distribution of hollow spheres are generally stronger. Since in the DHS we average over all possible inclusion fractions from 0 to 1, we always include spheres with extremely thin shells. These thin shells will, to a certain degree, simulate the optical properties of very small homogeneous particles, resulting in stronger spectral features. 
This can be understood by considering the particle as a combination of small volume elements, each interacting with the electromagnetic field like dipoles \citep[the Discrete Dipole Approximation;][]{1973ApJ...186..705P}. If these volume elements are close together, the interaction between the volume elements is important. However, if the volume elements are distributed over a larger area, as in a thin shell, the interactions between the volume elements are less important and the total interaction will be closer to the sum of the separate dipoles. In a thin shell the volume elements are distributed over a larger volume area than in a homogeneous spheroid with the same material volume, so the interaction of the light with the separate dipoles becomes more important.

As already mentioned, the 10 micron region is very important in astrophysics because of the possibility of ground based observations. Fig.~\ref{fig:Crystalline 10 micron} is a blowup of the 10 micron region. It shows that for the detailed shape of the spectrum, there are differences between the UDS and DHS results. In both sets of curves the trends as discussed above are still clear. Besides a change in the feature strengths and ratios, we also observe a shift in the wavelength positions of the resonances. For example the $11.3\,\mu$m feature shifts towards {$11.45-11.5\,\mu$m} (depending on the shape distribution used) when the grain size is increased from $r_V=0.1\,\mu$m to $r_V=1.0\,\mu$m. A similar shift from $11.3$ to $11.44\,\mu$m is observed when the chemical composition is changed from magnesium rich (forsterite, $y=1$) to iron rich (fayalite, $y=0$) \citep{Fabian, 2004ApJ...610L..49H}. Therefore, it is necessary to consider multiple resonances, at a large range of wavelengths, to disentangle the effects of particle size and chemical composition. Because we are looking at relatively short wavelengths, the spectral diagnostics of the emission resonances of forsterite are already reduced significantly when the volume equivalent radius of the particles is $\sim 2\,\mu$m. Therefore, the mineralogy that can be deduced from $10\,\mu$m spectroscopy is limited to particles smaller than a few micron. This has to be kept in mind when analyzing these spectra \citep[cf.][]{2004ApJ...610L..49H}.

\section{The degree of linear polarization}
\label{sec:The degree of linear polarization}

\begin{figure*}[!tbp]
\begin{center}
\resizebox{\hsize}{!}{\includegraphics{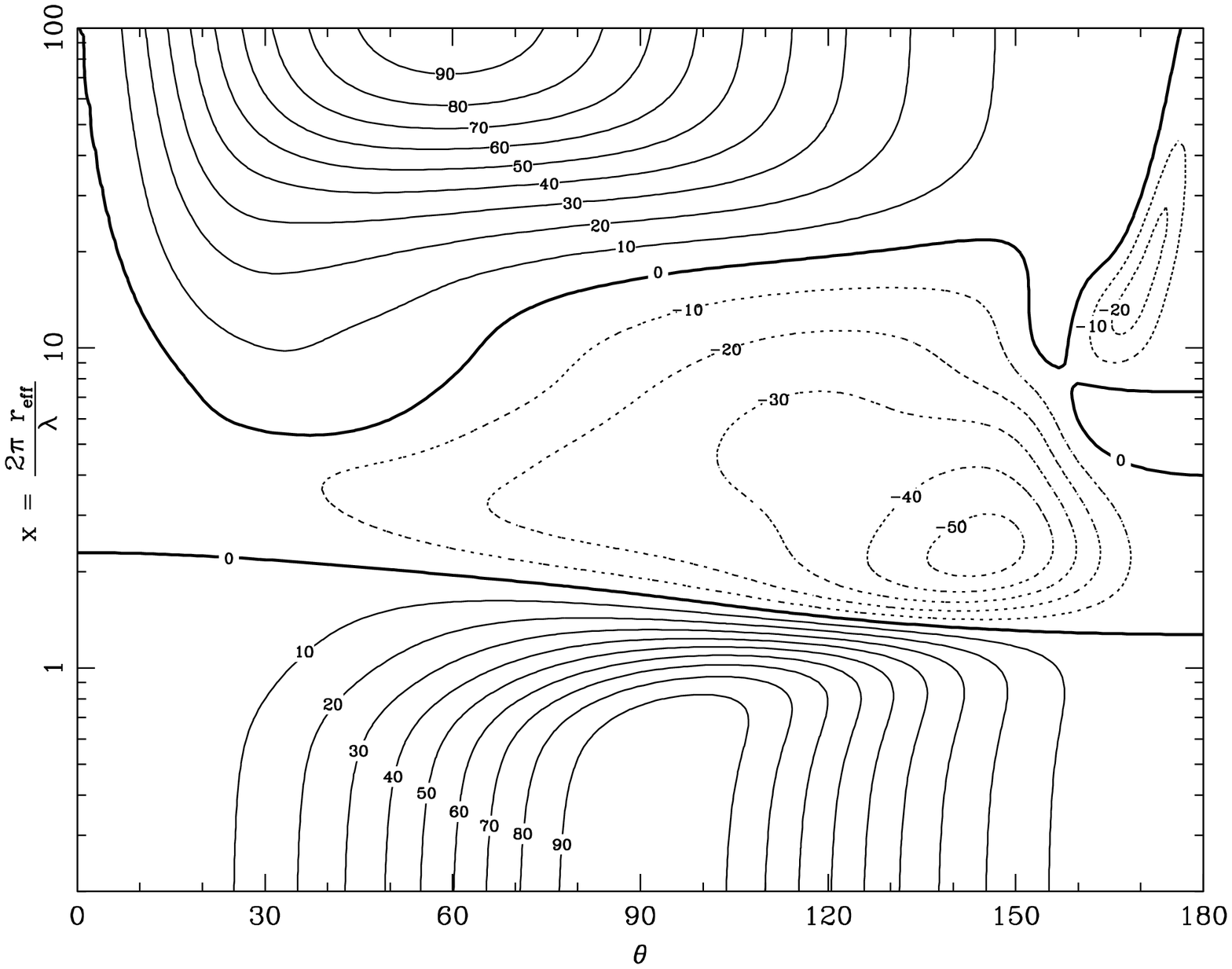}\includegraphics{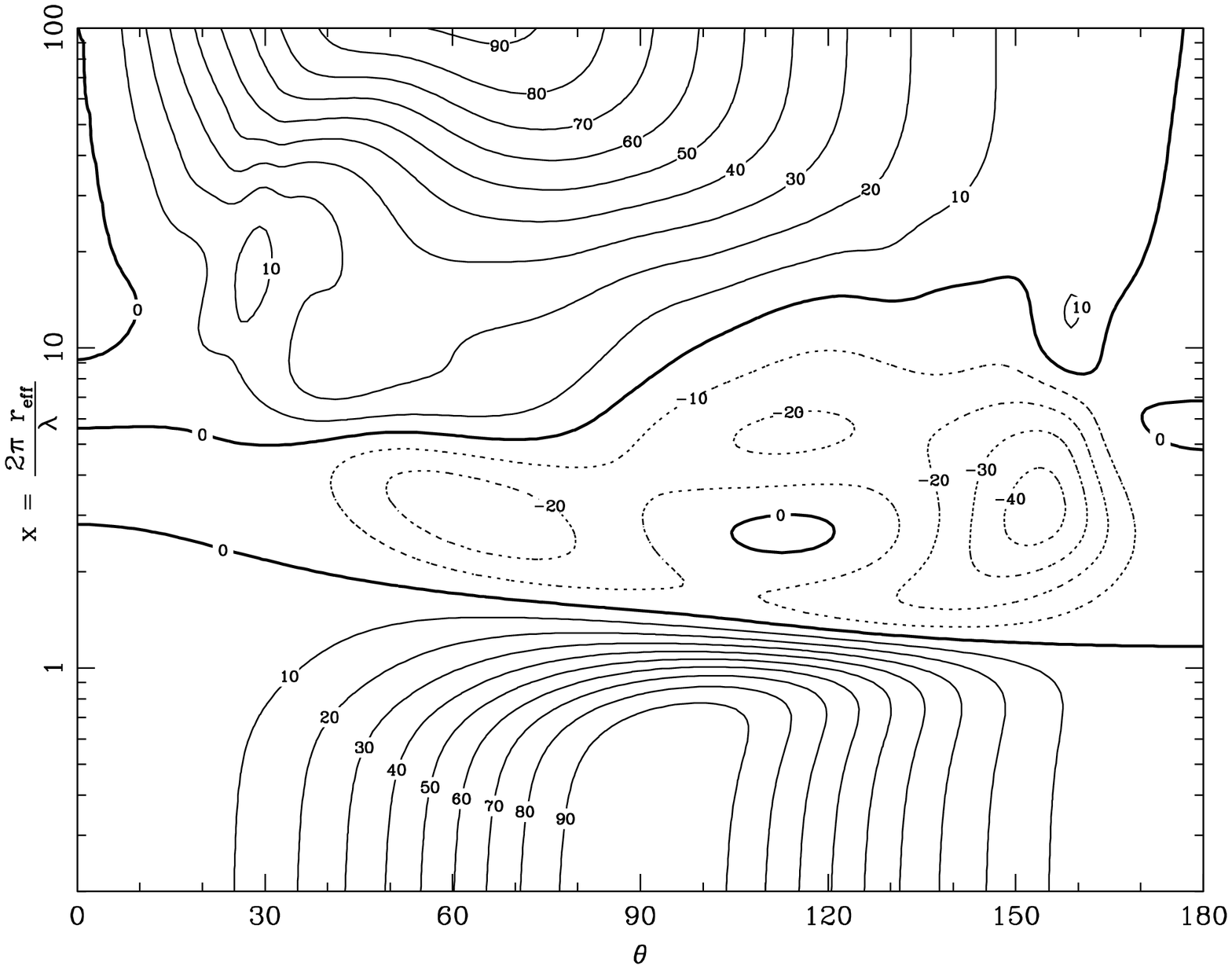}}\\
\resizebox{\hsize}{!}{\includegraphics{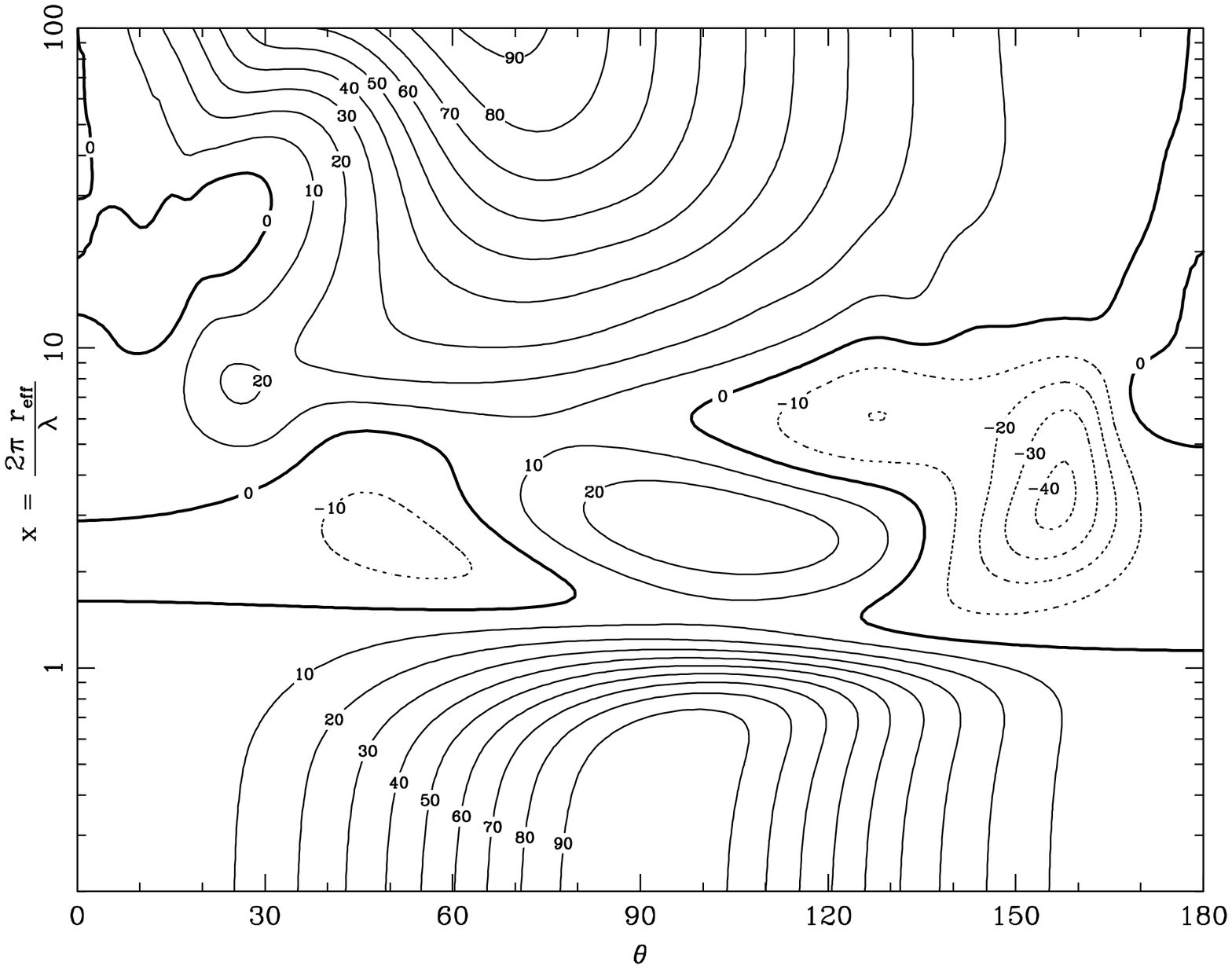}\includegraphics{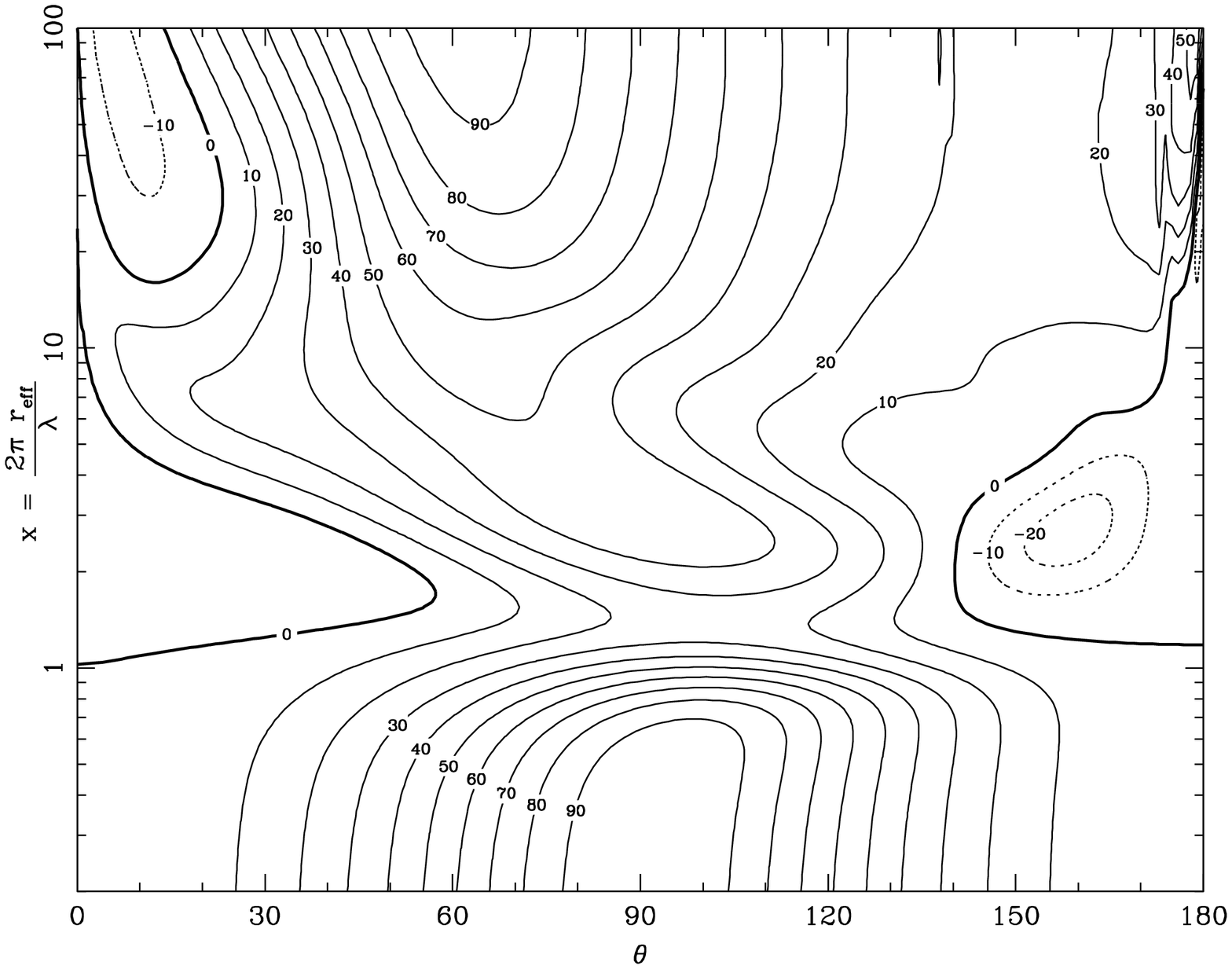}}
\end{center}
\caption{The degree of linear polarization of solid and hollow spheres as a function of the scattering angle and the effective size parameter for different values of $f_\mathrm{max}$. The refractive index of the material ${m=1.76+0.024i}$ which is typical for amorphous silicate with 30\% iron at {$\lambda=0.65\,\mu$m} \citep{1996A&A...311..291H}. Upper left ${f_\mathrm{max}=0}$ (homogeneous spheres), upper right ${f_\mathrm{max}=0.2}$, lower left ${f_\mathrm{max}=0.5}$ and lower right ${f_\mathrm{max}=1.0}$. A gamma size distribution with ${v_\mathrm{eff}=0.1}$ was used (see Eq.~(\ref{eq:gamma dist})). Solid and dotted curves represent the areas with positive and negative polarization respectively.}
\label{fig:Polarization Contours1}
\end{figure*}

\begin{figure*}[!tbp]
\begin{center}  
\resizebox{\hsize}{!}{\includegraphics{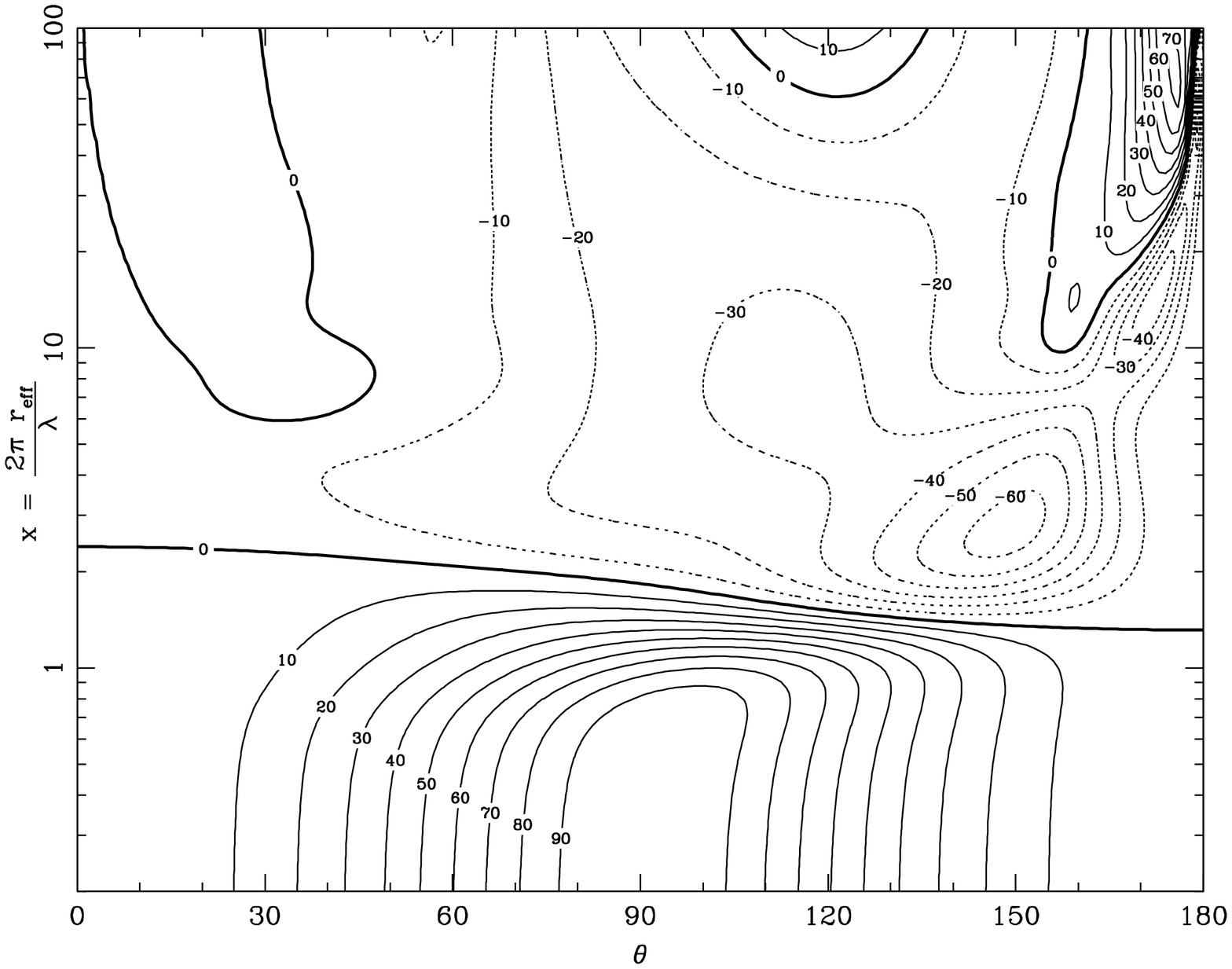}\includegraphics{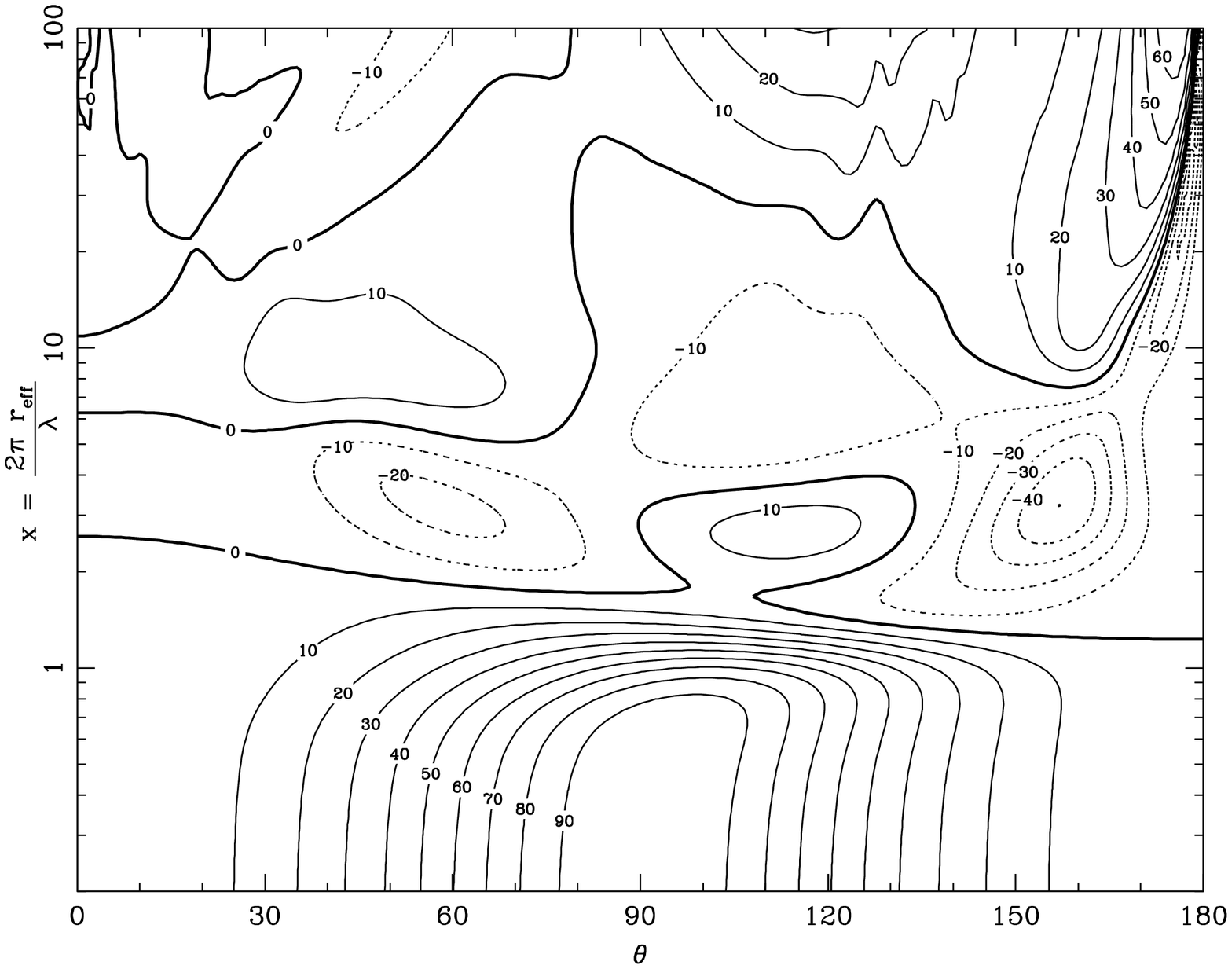}}\\
\resizebox{\hsize}{!}{\includegraphics{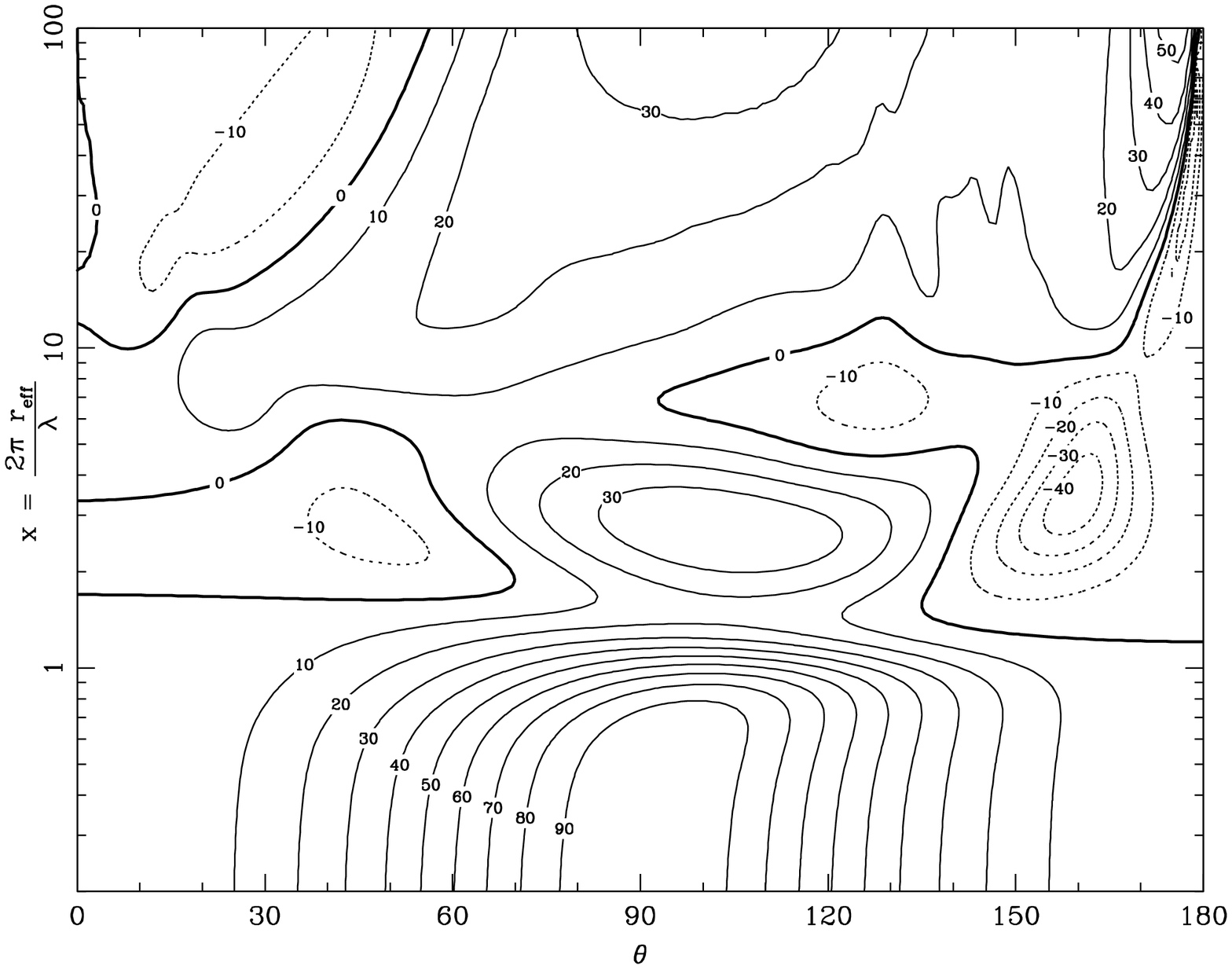}\includegraphics{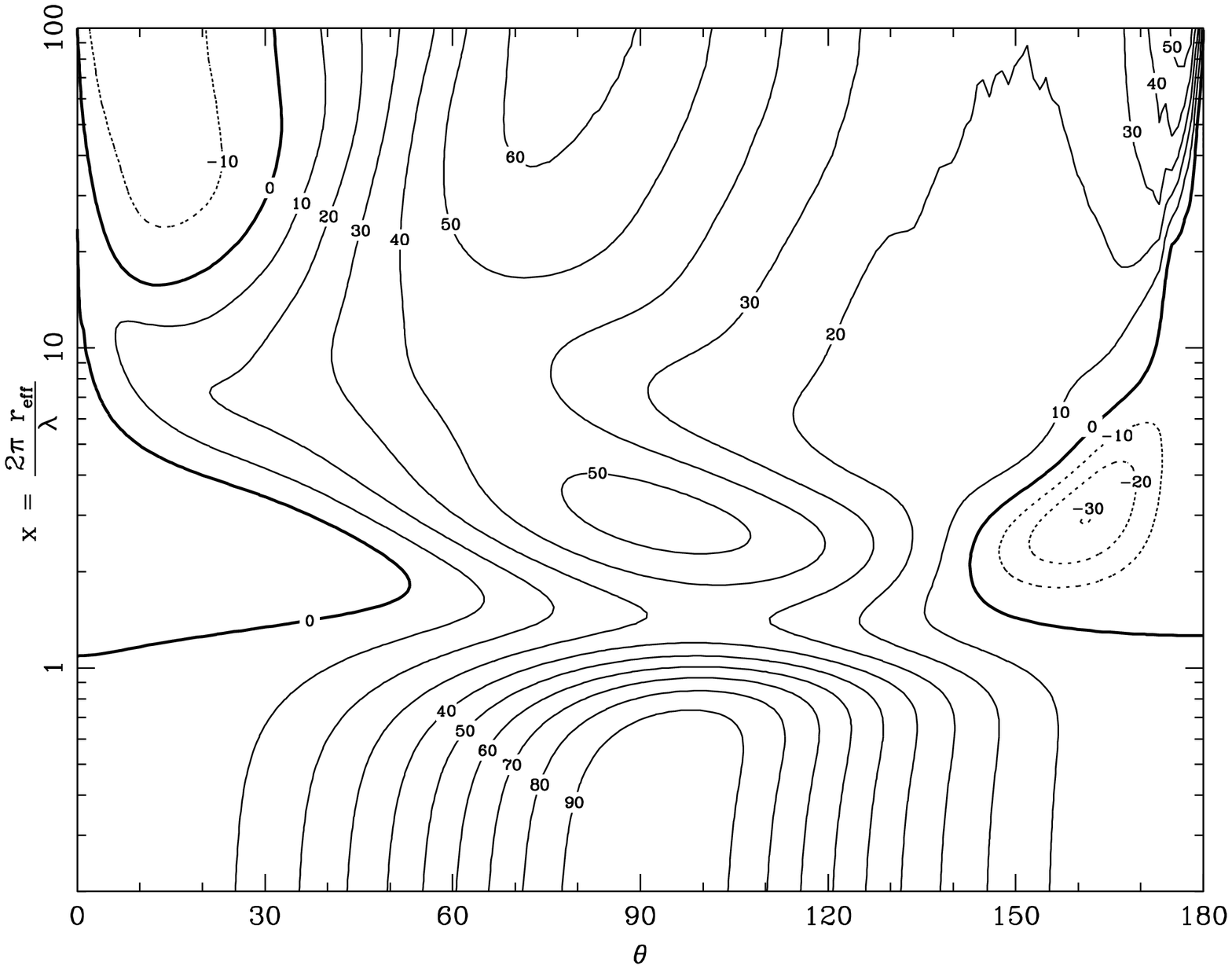}}
\end{center}
\caption{As Fig.~\ref{fig:Polarization Contours1} but for a refractive index of the material ${m=1.67+0.0001i}$ which is typical for iron poor amorphous olivine at {$\lambda=0.65\,\mu$m} \citep{1996A&A...311..291H}.}
\label{fig:Polarization Contours2}
\end{figure*}

Another area we wish to consider is the degree of linear polarization of the light scattered by dust grains in the visible part of the spectrum, which is a frequently used observable. For comets it is possible to obtain measurements for various angles of scattering by doing measurements at different orbital phases of the comet. For discussions on the polarization of light scattered by cometary dust and the possible implications for the grain properties see e.g. \citet{1999SSRv...90..163L} and \citet{2001JGR...10610113K}. The degree of linear polarization as a function of scattering angle can also be measured directly in the laboratory for collections of randomly oriented particles \citep[see e.g.][]{2003JQSRT..79..741H, Volten2001}. In most cases the shape of the linear polarization curve cannot be reproduced by using homogeneous spherical particles. Therefore it is very interesting to calculate the polarization properties of inhomogeneous or non-spherical particles. Many studies concerned the linear polarization of collections of spheroids or cylinders \citep{MishTravis2002, 2002JQSRT..74..167K, 2004JQSRT..85..231K}. Here, we will focus on calculations using the distribution of hollow spheres and compare our results with homogeneous spheres and laboratory measurements on irregular quartz particles.

Figs.~\ref{fig:Polarization Contours1} and \ref{fig:Polarization Contours2} show the degree of linear polarization of light scattered by collections of particles for incident unpolarized light as a function of scattering angle, $\theta$, and effective size parameter. Fig.~\ref{fig:Polarization Contours1} is for a refractive index typical for olivine with $y=0.5$, Fig.~\ref{fig:Polarization Contours2} is for a magnesium rich olivine with $y=1.0$. In order to reduce the effects of resonances a narrow gamma size distribution given by
\begin{equation}
\label{eq:gamma dist}
n(x)=c\cdot x^{(1-3v_\mathrm{eff})/v_\mathrm{eff}}\exp\left(-\frac{x}{v_\mathrm{eff}x_\mathrm{eff}}\right),
\end{equation}
was employed. In this equation $n(x)dx$ is the number of particles with size parameter $2\pi r_V/\lambda$ between $x$ and $x+dx$, $x_\mathrm{eff}$ is the effective size parameter, $v_\mathrm{eff}$ is the effective variance, which we take to be $0.1$, and $c$ is a normalization constant. For values of the effective size parameter smaller than approximately unity we see in Figs.~\ref{fig:Polarization Contours1} and \ref{fig:Polarization Contours2} the well known Rayleigh pattern with a maximum degree of linear polarization of $100$\% at $90^\circ$. For the homogeneous spheres and for the distribution of hollow spheres with ${f_\mathrm{max}=0.2}$ large areas in parameter space produce negative polarization. There are even values of the effective size parameter for which the linear polarization is negative over the entire range of scattering angles. For the most extreme distribution of hollow spheres (with ${f_\mathrm{max}=1}$) we see that the regions of negative polarization shift towards the very small and large scattering angles. This is consistent with a frequently observed negative polarization branch at large scattering angles (${\theta\approx 160-180^\circ}$) \citep{Volten2001, 2000A&A...360..777M, 1999SSRv...90..163L, 2001JGR...10610113K}. Also we note that the difference between the values obtained for olivine particles with $y=0.7$ and olivine particles with $y=1.0$ decreases when we increase $f_\mathrm{max}$. This suggests that for very irregular particles the refractive index of the material is of lesser importance. This agrees well with the observed similarity of the degree of linear polarization as a function of scattering angle for various mineral types \citep{Volten2001}.

The contours in Figs.~\ref{fig:Polarization Contours1} and \ref{fig:Polarization Contours2} for ${f_\mathrm{max}=1.0}$ show that the degree of linear polarization for this shape distribution is very high (up to 90\% at intermediate scattering angles) even for very large particles. The high degree of linear polarization seen here might be due to the fact that although the outer radius of the particles is large, the thickness of the shell is still very small compared to the wavelength, which causes a high degree of linear polarization (see also Sect. \ref{sec:Example: crystalline silicates}).

\subsection{Example: the polarization of small quartz particles}
\label{sec:Example: the polarization of small quartz particles}

\begin{figure*}[!tbp]
\resizebox{\hsize}{!}{\includegraphics{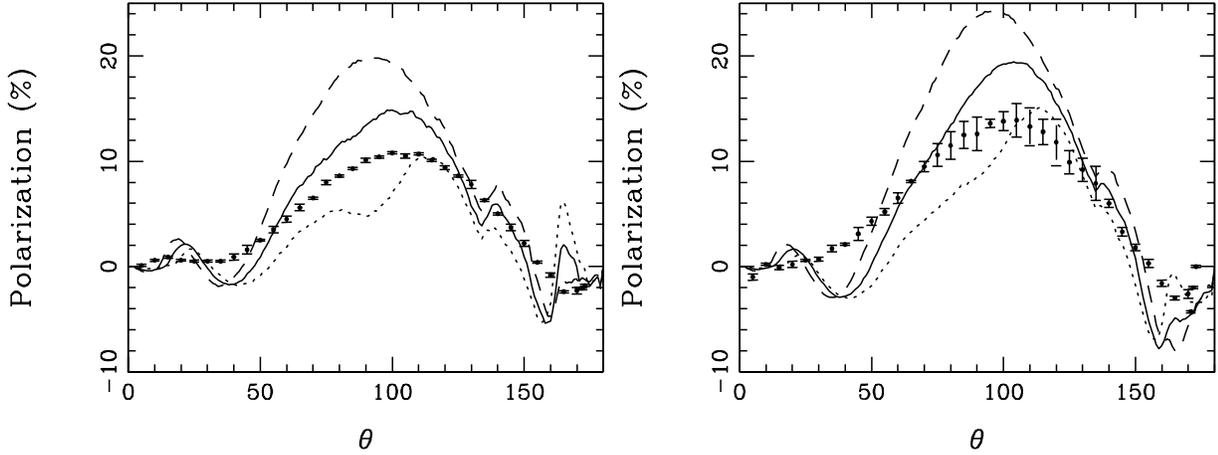}}
\caption{The degree of linear polarization of small quartz particles versus scattering angle, $\theta$. Shown are the measurements (points) and calculations (curves) using the measured size distribution and the distribution of hollow spheres with ${f_\mathrm{max}=0.3}$ (dotted curve), ${f_\mathrm{max}=0.4}$ (solid curve) and ${f_\mathrm{max}=0.5}$ (dashed curve). The left figure is for a wavelength $\lambda=422$\,nm and the right figure for $\lambda=633$\,nm. The refractive index is ${m=1.559+i\cdot10^{-8}}$ and ${m=1.542+i\cdot10^{-8}}$ for $\lambda=442$\,nm and $\lambda=633$\,nm, respectively \citep{1963aipa.book.....G}.}
\label{fig:pol real sdis}
\end{figure*}

Quartz, SiO$_2$, is one of the most abundant minerals in the Earth's crust. It is believed to be formed when amorphous silicates are annealed to form forsterite \citep{2000A&A...364..282F}. The refractive index of quartz in the visible part of the spectrum is very well known, which makes it an ideal test case for comparison between measurements and theory. \citet{Volten2001} measured the complete scattering matrix as a function of scattering angle in the range ${5-173^\circ}$ of an ensemble of small irregular quartz particles in random orientation at two wavelengths, $442$\,nm and $633$\,nm. The data of these measurements are available online together with the particle size distribution obtained from laser diffraction \citep{Volten2004}. From these scattering matrices we can obtain the degree of linear polarization for incident unpolarized light
\begin{equation}
P(\theta)=-\frac{F_{21}(\theta)}{F_{11}(\theta)},
\end{equation}
where $\theta$ is the scattering angle and $F_{nk}$ is the $n,k$th element of the $4\times 4$ scattering matrix \citep[for details see][]{Volten2004}. The laboratory data provide the element $F_{12}$. However, the symmetry of the system yields $F_{12}=F_{21}$. For comparison we also compute the degree of linear polarization using the distribution of hollow spheres together with the measured particle size distribution. For consistency with calculations elsewhere in the paper, we use volume equivalent radii to represent the size of the particles, although the measured size distribution is obtained for projected-surface-area equivalent spheres.

Fig.~\ref{fig:pol real sdis} shows calculated polarization curves for distributions of hollow spheres with ${f_\mathrm{max}=0.3,0.4}$ and $0.5$ together with the laboratory measurements. The results for ${f_\mathrm{max}=0.4}$ agree best with the measurements. This shape distribution reproduces the measured shape of the degree of linear polarization as a function of scattering angle reasonably well. The calculations show a feature at $\sim 170^\circ$ which may be caused by the fact that the particles in the hollow sphere distribution are still spherical in shape.

In most applications the real size distribution of the dust grains is not known. In these cases one would like to extract the size distribution from the measured polarization curve. To do this we try to minimize the difference between the measurements and the calculations by fitting the size distribution. 
To find the least squares solution we have to minimize $\chi^2$ with respect to the size distribution. The $\chi^2$ for this problem is defined by
\begin{equation}
\label{eq:reduced chi2}
\chi^2=\sum_j \left|\frac{-\frac{\sum_i w(r_i)~\left<F_{12}(\theta_j,r_i)\right>}{\sum_i w(r_i)~\left<F_{11}(\theta_j,r_i)\right>}-P_\mathrm{obs}(\theta_j)}{\sigma_j}\right|^2.
\end{equation}
In this equation $\left<F_{nk}(\theta_j,r_i)\right>$ is the $n,k$th element of the scattering matrix of an ensemble of randomly oriented particles with a certain shape distribution and a volume equivalent radius $r_i$ at scattering angle $\theta_j$, $w(r_i)$ is the value of the relative number size distribution at radius $r_i$ (these are the fitting parameters), $P_\mathrm{obs}(\theta_j)$ is the observed degree of linear polarization at scattering angle $\theta_j$ and $\sigma_j$ is the error in $P_\mathrm{obs}(\theta_j)$. The $r_i$ represent different values in the size distribution, and the $\theta_j$ represent the scattering angles at which measurements have been done. Unfortunately, Eq.~(\ref{eq:reduced chi2}) is not linear in the $w(r_i)$, which makes minimization of $\chi^2$ very difficult, especially in the case when we want to sample the size distribution with high resolution, which means we would have a large number of fitting parameters. However, we can linearize the equation by minimizing
\begin{equation}
\label{eq:linear step 1}
\chi'^2=\sum_j \left|\frac{-\sum_i  w(r_i)\left(\left<F_{12}(\theta_j,r_i)\right>-P_\mathrm{obs}(\theta_j)\left<F_{11}(\theta_j,r_i)\right>\right)}{\gamma_j~\sigma_j}\right|^2,
\end{equation}
where the $\gamma_j$ are scaling factors which can be different for each $\theta_j$. The right hand side of Eq.~(\ref{eq:linear step 1}) reduces to that of Eq.~(\ref{eq:reduced chi2}) when 
\begin{equation}
\label{eq:linear step 2}
\gamma_j=\sum_i w(r_i)~\left<F_{11}(\theta_j,r_i)\right>.
\end{equation}
We can now iteratively solve Eq.~(\ref{eq:reduced chi2}) and (\ref{eq:linear step 1}) in the following way. 
\begin{enumerate}
\item As a first estimate take $\gamma_j=1$ for all $j$. 
\item Solve Eq~(\ref{eq:linear step 1}) for the $w(r_i)$. 
\item The resulting values of $w(r_i)$ are used to generate new $\gamma_j$ from Eq.~(\ref{eq:linear step 2}).
\item Repeat steps 2 and 3 until convergence is reached, i.e. the $\gamma_j$ do not change during one iteration.
\end{enumerate}
In all cases that we considered convergence is reached within 15 iterations. This method is much faster than a nonlinear minimization algorithm and has the advantage that it is less sensitive to local minima.

\begin{figure*}[!tbp]
\resizebox{\hsize}{!}{\includegraphics{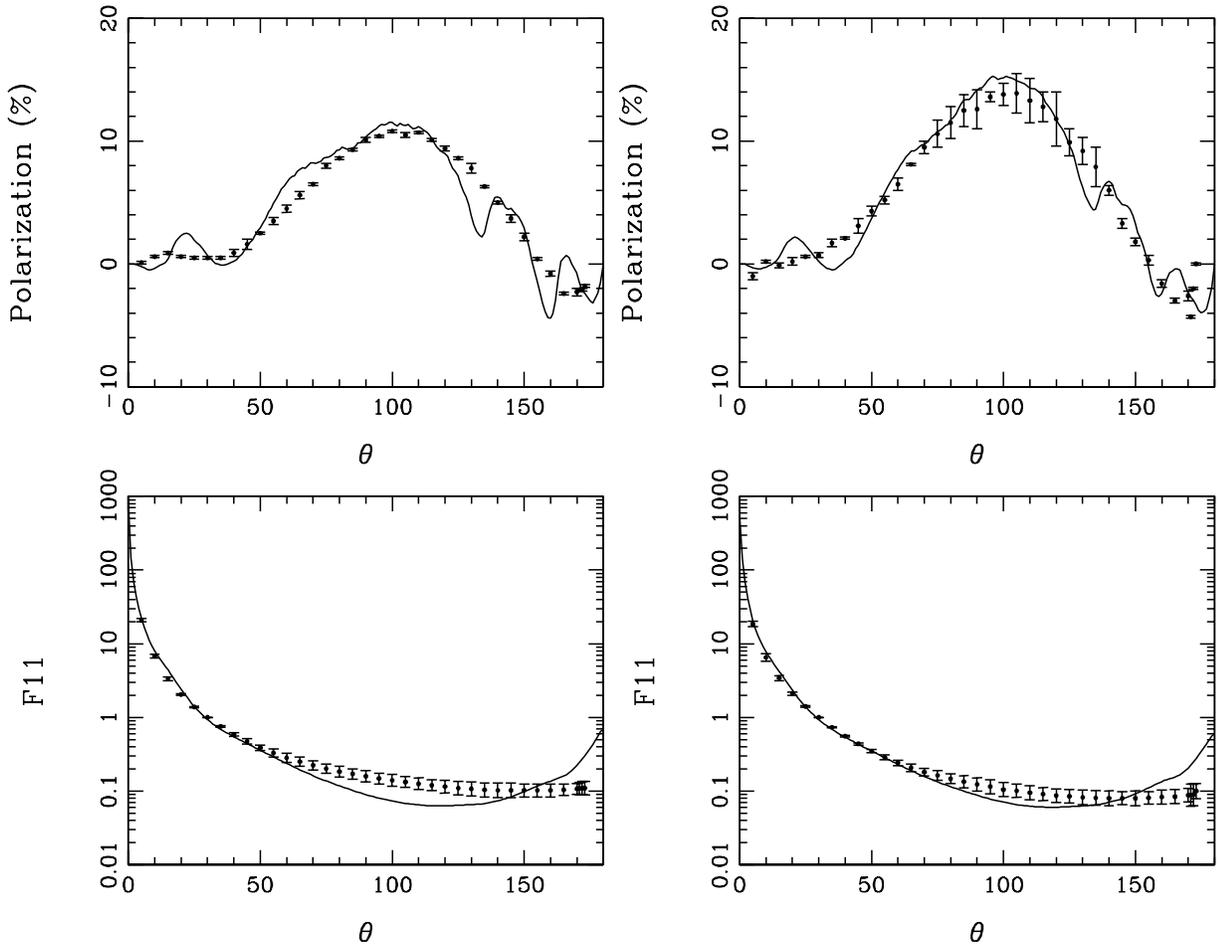}}
\caption{The degree of linear polarization (upper panels) and the phase function $F_{11}$ (lower panels) of small quartz particles versus scattering angle, $\theta$. All phase functions are normalized so that their values at $\theta=30^\circ$ equals unity. Shown are the measurements (points) and the best fit to the polarization measurements using the distribution of hollow spheres with ${f_\mathrm{max}=0.4}$ (solid curve). The left figures are for a wavelength $\lambda=422$\,nm and the right figures for $\lambda=633$\,nm. The refractive indices are as in Fig.~\ref{fig:pol real sdis}. The curves for the two different wavelengths have been fitted independently and the resulting size distributions are plotted in Fig.~\ref{fig:fitted sdis}. The phase functions have not been fitted but are calculated using the size distribution derived from the fit to the polarization measurements.}
\label{fig:pol fitted sdis}
\end{figure*}

If one would use the above scheme without further considerations, the solution for the $w(r_i)$ that is found will in general not be a smooth function of the $r_i$. Therefore, we add a regularization term to Eq.~(\ref{eq:linear step 1}) which is defined as
\begin{equation}
\label{eq:regularization}
\chi_R^2=\sum_i\left|w(r_i)-w(r_{i+1})\right|^2,
\end{equation}
and minimize ${\chi'^2+b~\chi_R^2}$ using the iterative scheme described above. The value of $b$ is chosen in such a way that the contributions from the two terms are balanced. For details on this regularization we refer to \citet{NumericalRecipes}.

We have to avoid finding solutions with $w(r_i)<0$ or the solution with $w(r_i)=0$ for all $i$. To do this we use a linear least squares fitting procedure with linear equality and inequality constraints. The constraints we use are ${\sum_i w(r_i)=1}$ and $w(r_i)>0$ for all $i$.

From the derived $w(r_i)$ we can construct the relative projected-surface-area distribution $S(\log r)$. This size distribution is defined in such a way that ${S(\log r)~\mathrm{d}\log r}$ gives the relative contribution to the total projected surface area of grains in the size range $\log r$ to ${\log r+\mathrm{d}\log r}$ \citep[see the appendix of][]{Volten2004}. We have to note here that we use volume equivalent radii of the particles while the measured size distribution is based on a projected-surface-area equivalent radius. This might introduce a small difference in the size distributions when the particles deviate strongly from homogeneous spheres. The differences here will be minimal.

\begin{figure}[!tbp]
\resizebox{\hsize}{!}{\includegraphics{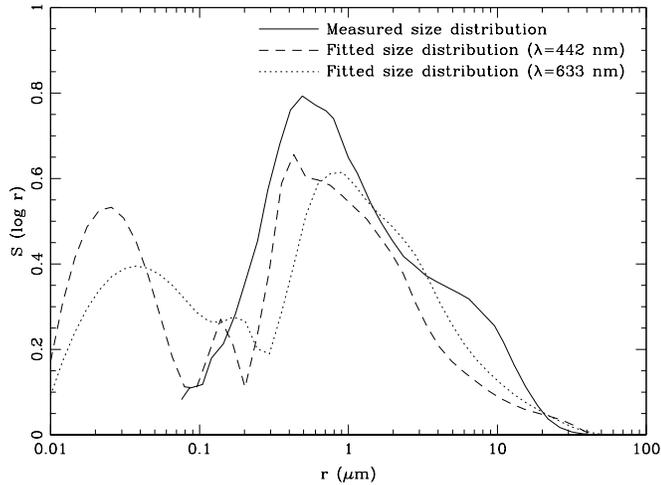}}
\caption{The size distribution, $S(\log r)$, of small quartz particles as measured by laser diffraction (solid curve) and as fitted from the linear polarization curve at $\lambda=442$\,nm (dashed curve) and at $\lambda=663$\,nm (dotted curve).}
\label{fig:fitted sdis}
\end{figure}

The best fit results are obtained when we take ${f_\mathrm{max}=0.4}$. The resulting polarization curves are shown in the upper panels of Fig.~\ref{fig:pol fitted sdis}. Plotted in the lower panels of this figure are the phase functions $F_{11}$ for the two different wavelengths. We see that the polarization curves are reproduced very well for both wavelengths. We have to note that for the other elements of the scattering matrix the agreement with the laboratory measurements is less convincing. For example depolarization and circular polarization effects cannot be explained using hollow spheres, because of the spherical symmetry of the particles. Also, as can be seen from Fig.~\ref{fig:pol fitted sdis}, the calculated phase functions display an increase at large scattering angles which is absent in the measurements.
The size distribution obtained by fitting the degree of linear polarization is plotted in Fig.~\ref{fig:fitted sdis} together with the measured size distribution using laser diffraction. We see that the grain sizes contributing most to the total projected-surface-area of all three size distributions match very well. In general, we find that an extra contribution of very small grains is required with respect to the measured size distribution in order to reproduce the measured polarization curve. The laser diffraction method used to measure the size distribution of the quartz particles in the laboratory is inaccurate for very small grains, which could be the reason why this small grain component is not found when this method is used. Another reason could be that the small grain component is used by the fitting procedure to simulate scattering by small scale structures in the particles itself. Considering the limited information used in the fitting procedure and the relatively simple grain shape distribution used, the similarity is remarkable. This is an encouraging result for the use of the distribution of hollow spheres in inverse problems to obtain the size distribution from the degree of linear polarization.
We have to stress here that having a reasonable estimate of the value of the refractive index is important for obtaining reliable results.

\section{Discussion}
\label{sec:Discussion}

In the previous sections it has been shown that the distribution of hollow spheres is a useful tool for calculating the optical properties of irregularly shaped particles. The question remains \emph{why} this is the case. It is clear that the particles in astrophysical environments or in the laboratory are not hollow spheres. Therefore, there must be a fundamental characteristic of the particles that determines their optical properties. This characteristic on the one hand has to be important for the scattering and absorption properties and on the other hand it has to be clearly different for homogeneous spheres and hollow or non-spherical particles.
One of the main things that discriminates a homogeneous sphere from other types of particles is its perfect symmetry. This perfect symmetry makes the interaction of an electromagnetic wave with this particle very sensitive to interference effects. This will determine to a large extent the optical properties. Only small deviations from a homogeneous sphere are required to disturb these interference effects. This suggests that the first disturbance in the perfect homogeneous sphere is very important and is not very shape-dependent.

\section{Conclusions}
\label{sec:Conclusions}

In this paper we examine the applicability of the statistical approach for computations of optical properties of ensembles of irregularly shaped particles in random orientation when a distribution of hollow spherical particles is used.

The range of particle sizes for which we did numerical computations covers a critical region if we wish to study the combined size and shape effects on the absorption, extinction and scattering cross sections as functions of the wavelength. For smaller particles we are in the Rayleigh regime, where the shapes of the spectra are independent of particle size. For particles larger than in the critical regime the effects of particle shape become negligible. We may thus conclude that, for all particle sizes, the emission and extinction spectra in the infrared computed by using the distribution of hollow spheres display the same trends and overall spectral shape as a uniform distribution of spheroids. These spectra are significantly different from those obtained using homogeneous spheres. This suggests that the shape of these spectra is, to a certain degree, not sensitive to the exact particle shape as long as the deviations from homogeneous spherical particles are large enough. This result is in agreement with the conclusion of \citet{Min2003b} for very small particles and is extended here to particles of arbitrary size.

The exact wavelength positions of observed spectral features are frequently used to identify the chemical composition of solid state particles. These wavelength positions also depend on particle size and shape. A reliable identification of solid state components can therefore only be made if the effects of both particle size and shape are taken into account. Calculations using the distribution of hollow spheres allow such a detailed analysis.

The degree of linear polarization for incident unpolarized light of hollow spheres with various volume fractions of the central vacuum inclusion was studied and compared to that of laboratory measurements at wavelengths of $442$ and $633$\,nm. Good agreement was found with laboratory measurements for small, irregular quartz particles for scattering angles between $5^\circ$ and $173^\circ$. It was also shown that the particle size distribution can be derived by fitting the calculated degree of polarization to the laboratory data. With this procedure we found an overabundance of small grains with respect to the measured size distribution. This can be due to the fact that the laser diffraction method used to measure the size distribution is inaccurate for small grains, or that the fitting procedure tries to simulate small scale structures that are present in the 'real' particles but absent in the hollow spheres. If the cause of the overabundance of small grains is the inaccuracy of laser diffraction when the particles are small, fitting the degree of linear polarization might be a good alternative for obtaining information on the particle size distribution.

Although our results imply that the optical properties of dust grains only contain limited information on the exact particle shape, by using the distribution of hollow spheres we have a more reliable diagnostic to derive the mineralogy, size and total mass of these dust grains than when only homogeneous grains and CDE are used.
We conclude that, because of its simplicity, low computational demand and success in reproducing the optical properties of irregular particles, the distribution of hollow spheres can be a powerful tool in studies of light scattering, absorption and emission by particles, and in particular in cases where large numbers of particle parameters need to be considered.

\begin{acknowledgements}
We are grateful to L.~B.~F.~M.~Waters and H.~Volten for enlightening conversations.
We would like to thank M.~I.~Mishchenko and O.~Mu{\~ n}oz for valuable comments on an earlier version of this paper.
\end{acknowledgements}

\end{document}